# Theoretical Analysis of Frequency-domain "single-shot" (FDSS) ultrafast spectroscopy. [a]

Ilya A. Shkrob[*][b] and Stanislas Pommeret [b,c]

[b] Chemistry Division , Argonne National Laboratory,  Argonne, IL 60439

[c] CEA/Saclay, DSM/DRECAM/SCM/URA 331 CNRS 91191 Gif-Sur-Yvette Cedex, France

**Abstract**

"Single-shot" ultrafast spectroscopy based on the frequency encoding of transient absorbance kinetics using chirped probe pulses is analyzed theoretically. FDSS has an advantage over pump-probe spectroscopy in a situation where the "noise" is dominated by amplitude variations of the signal. Unlike "single-shot" techniques based on spatial encoding of the kinetics, no a priori knowledge of the excitation profile of the pump is needed. FDSS spectroscopy can be used for many types of samples, liquid or solid, including those comparable in thickness to the wavelength of the probe light. Another advantage is that due to the interference of quasimonochromatic components of the chirped probe pulse, an oscillation pattern near the origin of the FDSS kinetics emerges. This interference pattern is unique and can be used to determine the complex dielectric function of the photogenerated species.

* To whom correspondence should be addressed: *Tel* 630-2529516, *FAX* 630-2524993, *e-mail:* shkrob@anl.gov.

b) Chemistry Division , Argonne National Laboratory, Argonne, IL 60439

c) CEA/Saclay, DSM/DRECAM/SCM/URA 331 CNRS 91191 Gif-Sur-Yvette Cedex, France

## I. INTRODUCTION

Most ultrafast laser spectroscopy experiments are done in a pump-probe fashion: a short probe pulse is delayed in time and then crossed in the sample with a pump pulse. The kinetics are obtained by plotting the transmission/reflection of the probe pulse as a function of the delay time, point by point. If the amplitude of the photoinduced signal varies from shot to shot or the sample is unstable (e.g., a high-speed jet), this variation translates into "noise" that is spread across the whole trace. The acquisition could be considerably shortened if there were a reliable method for "single-shot" sampling of the kinetics. Since the entire kinetic traces would modulate in the same way, shot-to-shot variations would not produce "noise" and the averaging efficiency would improve. Our interest in this "single-shot" methodology was instigated by the development of an ultrafast high energy electron source for radiation chemistry studies at Argonne.[1] Due to the high power load on the amplification optics, the repetition rate is rather low, < 5 Hz.[1] Owing to large shot-to-shot variations in the yield of electrons and this low repetition rate, pump-probe spectroscopy (PPS) is unsuited for this kind of electron source.

Two approaches for "single-shot" ultrafast spectroscopy have been previously suggested: (i) spatial encoding and (ii) frequency encoding. The former method was proposed by Rentzepis and coworkers [2,3] and further developed by Nelson and coworkers.[4,5,6] The method is similar in principle to the one used in frequency-resolved optical gating (FROG) devices.[7] In a typical setup,[4,5] cylindrically focussed pump and probe beams are crossed in the sample, and the transmitted probe and reference beams are imaged on a charge coupled device (CCD) array. Since the arrival time of the probe pulse at the sample continuously changes across the beam (on a sub-ps time scale), the pump and probe collide at different times and this "imprints" the kinetics on the axial profile of the transmitted light. The spread of the delay times can be increased further out to 10-15 ps by passing the probe (and/or the pump) beam through an echelon [3,6] or a supersonic beam.[8] Transient absorption (TA),[2,4,5,6] stimulated emission,[3] and grating [5] measurements on picosecond (ps) [2,3] and femtosecond (fs) [4,5] time scales have been carried out in this manner. A more sophisticated version of the same technique employs two crossed echelons to generate a two dimensional array of probe pulses separated by 30 fs in time that extends out to 10 ps. [6] While spatial encoding may work for some applications, there are several drawbacks that make this technique unsuitable for time resolved radiation chemistry:

First, it is almost impossible to extend the time window of the observation beyond the first 10-15 ps. Second, refraction of the probe and pump beams, by liquid jets and thermal lenses, is a serious concern, as it destroys the written pattern. For thin films, interference of the probe beam in the sample ruins the spatio-temporal correlation, and there are no examples of using the technique for samples thinner than 100 μm. Third, to obtain the spatial profile of the probe beam, one needs to know the spatial profile of photogenerated species, [4,5] as it is the product of these two profiles that is registered by the detector. To this end, a separate experiment with a calibrated sample is needed to obtain the pump beam profile. [4] If the photon order of the excitation is not known beforehand or the excitation is mixed order, the knowledge of such a profile is insufficient. Finally, it is impossible to focus the ultrashort pulse of relativistic electrons cylindrically or delay this pulse in time without substantial increase in its pulse duration, due to the energy spread. [1] From the engineering prospective, spatio-temporal encoding requires extremely high quality of the light beams and close placement of the detector to the sample, which makes it difficult to use this method in the confined space of a vacuum chamber.

The second technique, frequency encoding, is based on introducing frequency-dependent phase (i.e., chirp) in the probe pulse. [9,10] Frequency encoding has been used for "single-shot" temporal and spatio-temporal characterization of short light [11,12] and free electron [13] pulses, correlation spectroscopy, [14] holography, [15a] interferometry, [16,17] electro-optic measurements of THz pulses, [15b] etc. This encoding is implicit in the pump-probe spectroscopies that use chirped laser pulses, including transient absorption, [18,19] reflection, [19] interferometry, [17] and 4-wave mixing. [20] In 1992, Beddard et al. [10] suggested a chirped pulse TA spectroscopy in which the frequency encoding is used for "single-shot" *frequency domain* detection of kinetics, by spectral analysis of the transmitted probe light. In the following, this method is referred to as frequency-domain "single-shot" (FDSS) spectroscopy. Briefly, a linear chirp is introduced in the probe pulse. Since the blue light component of the chirped probe pulse arrives at the sample at a different delay time than the red light component, the absorption of these components by photogenerated species depends on their frequency, and the TA kinetics is imprinted in the *spectrum* of the transmitted probe pulse, with the "delay time" equal to the group delay at a given frequency. This technique can be used only for species whose TA spectra are either flat over the bandwidth of the probe pulse or known beforehand. Importantly, in FDSS spectroscopy, the probe light is not combined with the reference light at a spectrometer, as occurs in "single-shot" interferometry.[16,17] In the latter method, the kinetics are obtained by the Fourier transform of the resulting interferogram.[16] This makes FDSS kinetics simpler to obtain and analyze.

In the decade that passed after the experiments of Beddard et al.,[10] generation of ultrashort pulses using chirped-pulse amplification (CPA) with grating stretchers and compressors became routine.[9] The femtosecond pulses derived from a mode-locked Ti:sapphire laser provide the bandwidth that is optimum for FDSS measurements. Due to their high quality, these pulses can be negatively chirped using grating compressors that provide group velocity dispersion (GVD) up to a few ps$^2$.[9] As argued below, with the typical compressor designs that are used for CPA, "single-shot" kinetic measurements on the sub-nanosecond time scale are possible.

With these developments, FDSS has the potential to become a mainstream technique. In refs. 21 and 22 we demonstrated the advantages of FDSS for several photosystems, both solid and liquid. A variable-length grating compressor is used to change the observation window (by changing the GVD) over two decades in time. The proposed method bypasses most of the problems associated with spatial encoding and is easy to implement experimentally. In this paper, we give a detailed theoretical analysis of the FDSS technique as implemented in refs. 21 and 22. In Sec. II, an infinitely thin sample is considered. The more complex case of a sample of finite thickness (which may be comparable to the wavelength of the probe light) is examined in Appendix A. The requirements of the optics used for FDSS spectroscopy are discussed briefly in Sec. III; most of the technical details are given in Appendix B in the Supplement.[d] To save space, some figures (referred to in the text as, for example, Fig. 1S) are also placed in the Supplement.

## II. RESULTS AND DISCUSSION.

Our treatment is similar in approach to the previous analyses of chirped pulse pump-probe TA spectroscopy [18,19] and interferometry; [17] however, it departs from these in more than one respect. Taking $\Delta\omega = \omega - \omega_0$ as the frequency offset for the probe pulse, the Fourier transform $E_\omega$ of the electric field $E(t)$ for a chirped Gaussian pulse centered at $\omega_0$ is given by

$$E_\omega = E_{probe} \exp\left[-\Delta\omega^2 \tau_p^2/2 - i\phi(\omega)\right] \qquad (1)$$

where $E_{probe}$ is the time-integrated field and $\phi(\omega)$ is the frequency-dependent phase expanded as

$$\phi(\omega) \approx \frac{\phi''(\omega_0)}{2}\Delta\omega^2 + \frac{\phi'''(\omega_0)}{6}\Delta\omega^3 + ... \qquad (2)$$

where the first term corresponds to the group velocity dispersion (GVD) and the second and following terms correspond to higher order dispersion. [9,18] The zero and first order terms in $\Delta\omega$ are omitted, because the linear term simply shifts the origin of time. Neglecting higher than quadratic terms and introducing a stretch factor $s = \phi''(\omega_0)/\tau_p^2$ and a complex "width" of the pulse $\tau_c^2 = \tau_p^2(1+is)$, we obtain $E_\omega \propto \exp(-\Delta\omega^2 \tau_c^2/2)$. In the time domain, the oscillating field

$$E(t) = \frac{e^{-i\omega_0 t}}{\sqrt{2\pi}\tau_c} E_{probe} \exp\left(-t^2/2\tau_c^2\right) \qquad (3)$$

has a Gaussian envelope with a pulse width of $\tau_p\sqrt{1+s^2} \approx \tau_p s$ (for $s \gg 1$).

In the following, we will assume that the pump and the probe photons have vastly different energies and neglect coherent contributions to the TA signal, [23] as we are mainly interested in the dynamics that occur at delay times that are several times *longer* than the duration of the pump pulse. See ref. 18 for a more accurate derivation (for chirped pulse PPS) that takes into account both the sequential and Raman terms. [23] With our assumptions, the energy of the probe pulse absorbed in a thin sample of width *d* is given by the Joule law

$$\langle \mathcal{U} \rangle = \left\langle -E(t)\frac{d\Delta P(t)}{dt}\right\rangle \approx \frac{1}{2}\int_{-\infty}^{+\infty} d\omega \ \frac{\omega d}{c} \ \text{Im} \ E_\omega^* \ \Delta P_\omega, \tag{4}$$

where $\Delta P(t) = \Delta\varepsilon(t) \otimes E(t)$ is the polarization induced by the pump pulse, $\Delta\varepsilon(t) = \Delta\varepsilon'(t) + i\ \Delta\varepsilon''(t)$ is the photoinduced change in the complex dielectric function $\varepsilon$ of the sample and

$$\Delta P_\omega = \frac{1}{2\pi}\int_{-\infty}^{+\infty} dt\ e^{+i\omega t}\Delta\varepsilon(t)\otimes E(t) = \frac{1}{2\pi}\int_{-\infty}^{\infty}d\Omega\ E_\Omega\int_{-\infty}^{+\infty} dt\ e^{-i(\Omega-\omega)t}\Delta\varepsilon_\omega(t) \tag{5}$$

is the Fourier component of the third-order polarization. Since $|\Delta\omega| < 1/\tau_p \ll \omega$, the TA signal $S(\omega) = -\Delta T_\omega/T_\omega$ (where $T_\omega$ is the transmission of the sample at the frequency $\omega$) is given by

$$S(\omega) = 2\langle \mathcal{U} \rangle_\omega / |E_\omega|^2. \tag{6}$$

Taking into account finite resolution of the spectrometer, eq. (6) may be rewritten as

$$S(\omega) \ = \ -2\ \text{Re}\left[g(\omega)\otimes\frac{1}{t_\omega}\left(\frac{dt}{d\varepsilon}\right)_\omega\int_{-\infty}^{+\infty}d\Omega\ K_{\Omega-\omega}\ E_\Omega E_\omega^*\right]\Bigg/\left[g(\omega)\otimes|E_\omega|^2\right], \tag{7}$$

where $t_\omega$ is the Fresnel transmission coefficient for the probe light of frequency $\omega$ (this generalization of eq. (4) is justified in Appendix A),

$$g(\omega) = \left(\delta\sqrt{\pi}\right)^{-1}\exp\left(-\omega^2/\delta^2\right), \tag{8}$$

is the resolution function of the detector (for spectral power $\delta$) which is convoluted with the numerator and the denominator of eq. (7), and

$$K_{\Omega-\omega} = \frac{1}{2\pi}\int_{-\infty}^{+\infty} dt \ \Delta\varepsilon_\omega(t) \ \exp(-i(\Omega-\omega)t). \tag{9}$$

In the following, we will assume that the spectral profile of the photoinduced dielectric function $\Delta\varepsilon_\omega(t)$ is time independent. In this case, we can write $K_\mu = \Delta\varepsilon_\omega \ \cancel{K}_\mu$, where $\mu = \Omega - \omega$ and

$$\cancel{K}_\mu = \frac{1}{2\pi}\int_{-\infty}^{+\infty} dt \ \cancel{K}(t) \ \exp(-i\mu t) = \frac{\exp(i\mu T) \ \Pi_\mu \ \Gamma_\mu}{\Pi_{\mu=0}}. \tag{10}$$

The function $\cancel{K}(t) = \Pi(t+T) \otimes \Gamma(t)$ in eq. (10) gives the formation and decay kinetics of the photoinduced species, where $\Pi_\mu$ is the Fourier transform of the pump pulse $\Pi(t)$ (with the delay time $T$ given with respect to the center of the probe pulse) and $\Gamma_\mu$ is the Fourier component of the decay kinetics $\Gamma(t)$.

Eq. (7) was derived assuming that the transmission coefficient $t_\omega$ is a slow, nonoscillating function of $\omega$ in the spectral interval $\omega_0 \pm 1/\tau_p$; otherwise (e.g., for thin film samples), the more general eqs. (A18) and (A19) given in Appendix A must be used. For normal incidence of the probe light on a flat thin sample, $t_\omega^{-1}(dt/d\varepsilon)_\omega = i\omega d/c$. For a very thin wedge (considered below), $t_\omega^{-1}(dt/d\varepsilon)_\omega = i\omega d/c\sqrt{\varepsilon_\omega}$. For a Gaussian pump pulse, $\Pi(t) \propto |E_L(t)|^2 \propto \exp(-t^2/\tau_L^2)$ and exponential decay kinetics $\Gamma(t) = \exp(-\gamma t)$,

$$K_\mu = \Delta\varepsilon_\omega \ \exp(i\mu T - \mu^2\tau_L^2/4)/2\pi(\gamma + i\mu). \tag{11}$$

For infinite spectral resolution ($\delta = 0$), eq. (11) simplifies to

$$S(\omega) \approx \frac{2\omega d}{c} \text{Im} \ \Delta n_\omega \frac{1}{2\pi}\int_{-\infty}^{+\infty} d\mu \ \frac{\exp(i\mu T - \mu^2\tau_L^2/4)}{\gamma + i\mu} \frac{E_{\omega+\mu}}{E_\omega}, \tag{12}$$

where $\Delta n_\omega = \Delta \eta_\omega + i\, \Delta \kappa_\omega$ is the photoinduced change in the complex refraction index $n_\omega = \sqrt{\varepsilon_\omega}$ of the sample. If only the first term in eq. (2) is taken into account,

$$S(\omega) \approx \frac{2\omega d}{c} \operatorname{Im} \Delta n_\omega \Phi(\alpha, \beta(\Delta\omega), \gamma), \tag{13}$$

where we introduced a complex function

$$\Phi(\alpha, \beta, \gamma) = \frac{1}{2} \exp(\gamma^2 \alpha^2 - i\beta\gamma)\, \operatorname{erfc}\left(\gamma\alpha - \frac{i\beta}{2\alpha}\right), \tag{14}$$

with parameters $\alpha$ and $\beta$ given by equations

$$\alpha^2 = \frac{\tau_L^2}{4} + \frac{\tau_c^2}{2}\left(1 - \frac{\delta^2 \tau_c^2 / 2}{1 + \delta^2 \tau_p^2}\right), \tag{15}$$

$$\beta = -iT + \frac{\Delta\omega \tau_c^2}{1 + \delta^2 \tau_p^2}. \tag{16}$$

Eqs. (14), (15) and (16) are generalized to include finite spectral resolution, i.e., non-zero $\delta$. For $\delta = 0$, these expressions are similar to eq. (18) in ref. 18 that was obtained for chirped-pulse PPS by Kovalenko et al.[18] In the following, we will assume that $\delta$ is much smaller than the width $1/\tau_p$ of the probe pulse in the frequency domain (i.e., $\delta\tau_p \ll 1$) and use the group "delay time" $T_e$ defined as

$$T_e = -\operatorname{Im} \beta = T - \tau_p^2 s\, \Delta\omega. \tag{17}$$

For a sufficiently large frequency offset $\Delta\omega$ (long $T_e$), the error function in eq. (14) asymptotically approaches 2 and

$$S(\omega) \approx \frac{2\omega d}{c} \operatorname{Im} \Delta n_\omega \exp(\gamma^2\alpha^2 - i\beta\gamma) \propto \frac{2\omega d \Delta\kappa_\omega}{c} \exp(-\gamma T_e), \tag{18}$$

i.e., the signal $S(\omega)$ converges to exponential kinetics of TA, $\Delta\alpha_\omega(t) = \Delta\alpha_\omega\, \Gamma(t)$, where $\Delta\alpha_\omega = 2\omega d \Delta\kappa_\omega / c$, is the absorption coefficient at the probe wavelength $\omega$ and $t = T_e(\omega)$ (see Fig. 1). Formula (17) can be generalized for higher dispersion orders by noticing that for a sufficiently small frequency $\mu$ (that gives the largest contribution to the integral in eq. (12)),

$$E_{\omega+\mu}/E_\omega \approx \exp\left(-\mu\{\Delta\omega\tau_p + i\phi'(\omega)\} - \mu^2/2\{\tau_p^2 + i\phi''(\omega)\}\right). \tag{19}$$

Substituting this formula into eq. (12), the integral can be reduced to function (14) with

$$\alpha^2 = \frac{\tau_L^2}{4} + \frac{\tau_c^2}{2} + \frac{\phi''(\omega)}{2} \quad \text{and} \quad i\beta = T + i\Delta\omega\tau_p - \phi'(\omega), \tag{20}$$

so that $T_e$ is equal to the group delay at frequency $\omega$ [18]

$$T_e \equiv -\text{Im}\,\beta = T - \phi'(\omega) \approx T - \phi''(\omega_0)\Delta\omega - \frac{\phi'''(\omega_0)}{2}\Delta\omega^2 - \ldots. \tag{21}$$

Since both parameters $\alpha$ and $\beta$ are complex functions of $\omega$, $S(\omega)$ oscillates strongly near the kinetics origin, where $T_e = 0$ (the real and complex parts of function $\Phi$ are shown in Figs. 1 and 2). It was these oscillations that were observed by Beddard et al. [10] The oscillations are stronger for smaller $\tau_L$ and $\delta$, whereas longer pump pulses and lower spectral resolution damp these oscillations (see below). To characterize the oscillation pattern, we will assume that $T = \tau_L = 0$ and $|s| \gg 1$. Then, for $\gamma = 0$ (a step-like kinetics),

$$S(\omega) \propto \text{Im}\,\exp(i\phi_\varepsilon)\,\text{erfc}\left((1-i)/2\,\Delta\omega\tau_{GVD}\right), \tag{22}$$

where $\phi_\varepsilon$ is the phase of $\Delta n_\omega$, $\tan\phi_\varepsilon = \Delta\kappa_\omega/\Delta\eta_\omega$, and $\tau_{GVD} = \sqrt{|\phi''(\omega_0)|}$. Differentiating both sides of eq. (22) with respect to $\omega$ yields

$$\partial S(\omega)/\partial\omega \propto \sin\left(\Delta\omega^2\tau_{GVD}^2/2 - \pi/4 + \phi_\varepsilon\right). \tag{23}$$

Thus, the stationary points of $S(\omega)$ are symmetric about the origin at $T_e = 0$. Introducing the time interval $\Delta T_e^{(n)}$ between the $n$-th pair of these points in the group delay, we obtain

$$\Delta T_e^{(n)} = \tau_{GVD} \left\{ 2 \left( \pi - 4\phi_\varepsilon + 4\pi n \right) \right\}^{1/2}, \tag{24}$$

where the index $n = 0,1,...$ must be sufficiently large so that the square root exists. For the first pair of these stationary points,

$$\Delta T_e^{(1)} = \tau_{GVD} \sqrt{6\pi} \quad for \quad \phi_\varepsilon = \pi/2 \tag{25}$$

and

$$\Delta T_e^{(0)} = \tau_{GVD} \sqrt{2\pi} \quad for \quad \phi_\varepsilon = 0. \tag{26}$$

Thus, the oscillation patterns for photoinduced absorption ($\phi_\varepsilon = \pi/2$) and (nonlinear) refraction ($\phi_\varepsilon = 0$) are quite different, both in their phase (eqs. (22)) and in the oscillation frequency (eq. (24) and Figs. 1 and 2). Since the error function in eq. (14) becomes real for long group delay times $T_e$, for $\phi_\varepsilon = 0$ the nonlinear refraction contribution $S(\omega)$ tends to zero for $|T - T_e| \gg \tau_{GVD}$ (the upper traces in Figs. 1 and 2). Thus, $S(\omega)$ is sensitive to nonlinear refraction only near the kinetics origin, where $T_e = 0$. To see how the finite spectral resolution causes damping of $S(\omega)$, consider the case when the decay kinetics $\Gamma(t)$ is so fast that one may let $\gamma \to \infty$. In this case, we obtain

$$S(\omega) \approx \mathrm{Im} \, \Delta\varepsilon_\omega \left(2\sqrt{\pi}\gamma\alpha\right)^{-1} \exp\left(\beta^2/4\alpha^2\right). \tag{27}$$

For $T = 0$ and $s \gg 1$

$$\exp\left(\beta^2/4\alpha^2\right) \approx \exp\left(-T_e^2 / \left\{\tau_L^2 + \delta^2 \tau_{GVD}^4 - 2i\tau_{GVD}^2\right\}\right) \tag{28}$$

and $S(\omega)$ is damped as $\exp\left(-\left[T_e/T_{damp}\right]^2\right)$, where $T_{damp}$ is given by the real part of eq. (28). For $\tau_L \approx 0$ and $\delta\tau_{GVD} < 1$, $T_{damp} \approx 2/\delta$. Thus, for $\delta=1$ cm$^{-1}$, the oscillations in the $T_e$ domain are damped in 10.6 ps (Fig. 3 illustrates the oscillation damping for $\delta=2, 5,$ and 10 cm$^{-1}$). For $\delta \approx 0$, $T_{damp}^2 = \tau_L^2 + \left(2\tau_{GVD}^2/\tau_L\right)^2$, i.e., for $\tau_L > \tau_{GVD}/2$, at most one oscillation would survive (see Fig. 3).

Using eqs. (13) to (16), one can find the optimum set of parameters needed to sample an exponential kinetics with a given time constant $\gamma^{-1}$:

$$\Delta\omega_{max} \approx 1/\tau_p, \ s \approx \pm 2/\gamma\tau_p, \ and \ T \approx \pm 1.25/\gamma. \quad (29)$$

The first equation specifies the optimum spectral range $(\omega_0 - \Delta\omega_{max}, \omega_0 + \Delta\omega_{max})$ for which $|E_\omega|^2$ is reasonably large at the extremes of this range. The next two equations specify the optimum stretch factor $s$ of the probe pulse and the pump delay time $T$ chosen to take the maximum advantage of the spectral range (where $s > 0$ for a stretched pulse and $s < 0$ for a compressed pulse). If only the kinetic profile of TA for $|T - T_e| \geq (2-5)\tau_{GVD}$ is of interest, the width $\tau_L$ of the pump pulse can be chosen so that $\tau_L \geq (0.5 - 1)\tau_{GVD}$; in this case the oscillations are almost completely eliminated (Fig. 3). To the same end, one can reduce the spectral resolution $\delta$ so that $\delta\tau_{GVD} > 1$ (Fig. 3). If the oscillation pattern of $S(\omega)$ is of interest (e.g., for the measurement of $\Delta n_\omega$), one needs to minimize both $\tau_L$ and $\delta$.

Fig. 4 shows the real part of function $\Phi$ (eq. (14)) for $\tau_p = 20\,fs$ and several stretch factors $s$. For picosecond kinetics ($s=100$), 30-50% of the time window is taken by the oscillation pattern; for larger GVD, this fraction decreases as $|s|^{-1/2}$. The observed trends agree with the qualitative analysis given above and the experimental results in refs. 21 and 22. Fig. 5 shows the progression of signals $S(\omega)$ as a function of the phase $\phi_\varepsilon$ for $\Delta n_\omega = const(\omega)$. Observe how the oscillation pattern near the kinetic origin changes with the phase; this dependence may be used to estimate $\phi_\varepsilon$ from the general shape of the oscillatory pattern (see ref. 21 for experimental demonstration of such a measurement). For $\Delta n_\omega$ that spectrally evolves in time this pattern reflects the average phase over a period of time $\sim \tau_{GVD}$. By changing GVD, one can obtain the time dependence of this average phase.

### III. EXPERIMENTAL CONSIDERATIONS.

As compared to CPA, FDSS sets very modest constraints on the compressor or stretcher used to chirp the probe pulse. In the usual CPA scheme,[9] a short pulse from the oscillator is stretched, amplified, and then compressed, so that the positive chirp added by the stretcher and the amplifier optics is cancelled by the negative chirp added by the compressor. Therefore, to obtain a transform limited, ultrashort pulse, this scheme requires compensation of third (TOD) and higher order dispersion. In the FDSS, TOD and the higher orders are unimportant because for the standard beam geometry, a grating compressor/stretcher cannot introduce sufficiently large TOD to cause notable distortion of the kinetics. For a single round-trip between the pair of parallel gratings, the GVD is given by [9]

$$\phi''(\omega_0) = - \left(\lambda^3 L_g / \pi c^2 d_g^2\right) \left[1 - \left(\lambda/d_g - \sin\theta\right)^2\right]^{-3/2}, \qquad (30)$$

and the TOD (that is, $\phi'''(\omega_0)$) can be found from the equation

$$\xi_3 = \omega_0 \phi'''(\omega_0) / 3\phi''(\omega_0) = -\left(\sin\theta \ \lambda/d_g + \cos^2\theta\right) / \left[1 - \left(\lambda/d_g - \sin\theta\right)^2\right], \qquad (31)$$

where $\lambda = 2\pi c/\omega_0$ is the wavelength of the probe light at the center, $\theta$ is the angle of incidence on the grating (chosen to be as close as possible to the Littrow angle, $\sin\theta_L = \lambda/2d_g$), $L_g$ is the slanted distance between the gratings, and $d_g = d_g(m)$ is the ratio of the groove spacing and the diffraction order $m$. For a stretcher, GVD has the opposite sign. Note that both GVD and $\xi_3$ are maximum at $\theta_L$, where $\xi_3 \approx -\left(1 + \lambda^2/4d_g^2\right)/\left(1 - \lambda^2/4d_g^2\right)$ is close to -1. We can rewrite eq. (21) as

$$T_e \approx T - \Delta\omega \; \phi''(\omega_0) \left[1 \;+\; 3\xi_3/2 \; (\Delta\omega/\omega_0)\right]. \tag{32}$$

Since $\Delta\omega\tau_p < 1$ in the optimum spectral range (eq. (28)), the correction to the second term of eq. (32) does not exceed $(1-2) \times (\omega_0\tau_p)^{-1}$ - which is at most a few per cent. Fig. 1S(a) shows a simulation of $S(\omega)$ as a function of $T_e$ defined by eq. (17) for $\xi_3 = -1.6$ (that corresponds to the Littrow angle dispersion of 800 nm probe light by a 1200 $g$/mm grating in the first diffraction order). While the TOD shifts the kinetic origin, the change in the kinetic profile is very slight. Furthermore, this small change can be completely eliminated by plotting the kinetics as a function of the group delay $T_e$ defined by eq. (32), as shown in Fig. 1S(b).

Another important point is that the probe beam dispersed on the grating can be clipped by the grating edges. In CPA, the spectral wings cannot be clipped because the high-frequency components must be added to prevent oscillations of the compressed pulse in the time domain.[9] By contrast, in FDSS, the kinetics are imprinted in the *central* section of the probe spectrum and Fourier components outside this section simply do not matter. This point is demonstrated by numerical simulations in Appendix B.1 in the Supplement.[d] According to this analysis, for the same stretch factor, the gratings in the FDSS setup can be placed 3-5 times further away than the gratings in the equivalent CPA design. This calculation suggests that with the standard CPA optics, one can readily sample the kinetics over 300-500 ps.

Since the spectral range and resolution are fixed by the choice of the monochromator and detector, to obtain the kinetics on different time scales, GVD must be tuned over the widest possible range. To this end, a variable-length grating compressor is best suited.[21,22] In the FDSS setup described in ref. 21, the probe pulse is taken from the 800 nm light, positively chirped using a fixed-length $L_g$=40 cm grating stretcher, and then negatively chirped using a variable-length compressor. A separate fixed-length compressor is used to generate a transform-limited, amplified 800 nm pulse that provides the pump light. This scheme allows for continuous change in the probe GVD between 0 and -1.6 ps$^2$. Inexpensive gratings can be used, as large power losses can be tolerated due to high intensity of the probe light. For the same reason, higher diffraction orders $m$ can be used to increase GVD without changing $L_g$ (eq. (30)).

The greatest inconvenience presented by the FDSS technique is that the probe color cannot be easily changed. A strategy of overcoming this deficiency, by using compressor gratings in higher order, is given in Appendix B.2 in the Supplement. Another practical concern, nonideality of the probe pulse (i.e., the deviation of its power spectrum from an ideal Gaussian profile), turns out to be of relatively little significance for FDSS spectroscopy, as discussed in Appendix B.3 in the Supplement.

## IV. CONCLUSION.

Ultrafast frequency-encoded "single-shot" transient absorption spectroscopy (FDSS) is analyzed theoretically within the framework of a generalized treatment given in Sec. II. It is shown that interference of quasimonochromatic components of the chirped probe pulse introduces an oscillatory pattern near the kinetic origin with a time period comparable to the square root of the GVD ($\tau_{GVD}$). The fraction of FDSS kinetics that is modulated by these oscillations decreases as $|s|^{-1/2}$ with the increase in the absolute stretch/compression factor $s$ of the probe pulse. For long-lived FDSS signals, this oscillatory pattern can be used to estimate the phase of the complex dielectric function.

Due to the practical importance of thin-film semiconductors, a more general formulation of our approach for such samples is given in Appendix A; the correctness of this general approach is demonstrated experimentally in ref. 22. In refs. 21 and 22, the theoretical results of Sec. II and Appendix A are used to interpret experimental FDSS kinetics for photoinduced charge separation in molecular liquids and semiconductor solids.

The experimental concerns regarding the use of variable length grating compressor for chirping the probe pulse for FDSS spectroscopy are addressed in Sec. III. As shown there, standard optics used for chirped pulse amplification can be used to extend the time window of FDSS to 300-500 ps. These FDSS kinetics can be spliced together to obtain even longer time profiles. [21] In this respect, and in the ability to vary the observation window over two decades in time, FDSS is superior to the methods based on spatio-temporal encoding (Sec. I). The only strict requirement imposed by the technique on the probe pulse is that the spectrum should be reasonably close to a Gaussian. Still, a "ripple" pattern with peak-to-peak amplitude of 10-20% can be tolerated (Appendix B.3). With regard to the time resolution, the main shortcoming of FDSS spectroscopy is that the kinetics shorter than a few picoseconds exhibit strong oscillations and may be of limited use without analysis. For applications where amplitude variations of the kinetics are large, absorption spectra of photogenerated species are broad and featureless, repetition rates are low, and subpicosecond resolution is unimportant, FDSS has a clear advantage over pump-probe spectroscopy and many other "single-shot" spectroscopies.

## V. ACKNOWLEDGMENT

This work was performed under the auspices of the Office of Basic Energy Sciences, Division of Chemical Science, US-DOE under contract No. W-31-109-ENG-38. SP acknowledges the support of the DGA through the contract number DSP/01-60-056.

## Appendix A. Arbitrary thickness sample on a clear substrate.

In this Appendix, a general case of the probe light propagating through a thin sample is examined. Without loss of generality, we will assume that the sample is a flat layer of thickness $d$ with a complex dielectric function $\varepsilon_\omega = n_\omega^2$ on a thick, non-absorbing substrate with index of refraction $n_s$. The propagation of light through the sample in the direction normal to the surface is given by a wave equation

$$c^2 \frac{\partial^2 E(t,z)}{\partial z^2} = \{\varepsilon + \Delta\varepsilon(t)\} \otimes \frac{\partial^2 E(t,z)}{\partial t^2}, \tag{A1}$$

where $z$ is the depth of the sample ($z=0$ for the air/sample interface and $z=d$ for the sample/substrate interface) and $\Delta\varepsilon(t)$ is the perturbation of the complex dielectric function by the pump light. We will assume that this perturbation is independent of $z$, i.e., the pump light is absorbed homogeneously throughout the sample (if the energy deposition is $z$-dependent, even the PPS experiment becomes difficult to analyze rigorously).[24] To solve eq. (A1) we will use first order perturbation theory and assume that

$$E(t,z) \approx E^0(t,z) + \Delta E(t,z), \tag{A2}$$

where the zero and the first order terms satisfy the equations

$$c^2 \frac{\partial^2 E^0(t,z)}{\partial z^2} = \varepsilon \frac{\partial^2 E^0(t,z)}{\partial t^2}, \tag{A3}$$

$$c^2 \frac{\partial^2 \Delta E(t,z)}{\partial z^2} - \varepsilon \frac{\partial^2 \Delta E(t,z)}{\partial t^2} = \Delta\varepsilon(t) \otimes \frac{\partial^2 E^0(t,z)}{\partial t^2}. \tag{A4}$$

The first of these equations has the solution

$$E^0(t,z) = \int_{-\infty}^{\infty} d\omega \ e^{-i\omega t}\left(E_\omega^{0,+}e^{+ik_\omega z} + E_\omega^{0,-}e^{-ik_\omega z}\right), \tag{A5}$$

where $k_\omega = \omega n_\omega/c$ is the wave vector of the light with frequency $\omega$ and $n_\omega = \eta_\omega + i\kappa_\omega$ is the complex refraction index. Coefficients $E_\omega^{0,\pm}$ in eq. (A5) may be found from the continuity of the electric field $E(t,z)$ and its derivative $\partial E(t,z)/\partial z$ with respect to the co-ordinate $z$ at $z=0$ and $z=d$:

$$\begin{aligned}(1+r_\omega)E_\omega^i &= E_\omega^{0,+} + E_\omega^{0,-}, \quad (1-r_\omega)\ E_\omega^i = n_\omega(E_\omega^{0,+} - E_\omega^{0,-}),\\ t_\omega E_\omega^i &= E_\omega^{0,+}e^{i\delta_\omega} + E_\omega^{0,-}e^{-i\delta_\omega}, \quad n_s\ t_\omega E_\omega^i = n_\omega(E_\omega^{0,+}e^{i\delta_\omega} - E_\omega^{0,-}e^{-i\delta_\omega}),\end{aligned} \tag{A6}$$

where $(r,t)_\omega$ are the reflection and transmission (Fresnel) coefficients, respectively, $\delta_\omega = k_\omega d$ is the complex phase acquired by the Fourier component with frequency $\omega$ as it propagates through the sample, and $E_\omega^i$ is the amplitude of the incident light. Eqs. (A6) can be solved to obtain the Fresnel coefficients: [24]

$$t_\omega = 2n_\omega/D_\omega \quad and \quad r_\omega = \{\cos\delta_\omega\ n_\omega(1-n_s) + i\ \sin\delta_\omega\ (n_\omega^2 - n_s)\}/D_\omega, \tag{A7}$$

$$D_\omega = \cos\delta_\omega\ n_\omega(1+n_s) - i\ \sin\delta_\omega\ (n_\omega^2 + n_s). \tag{A8}$$

Neglecting the reflections from the back surface of the substrate, the transmission coefficient $T_\omega$ for the light passing through the sample on the thick substrate is equal to $t_s \times t_\omega$, where $t_s = 2n_s/(n_s+1)$ is the transmission coefficient for the substrate. The photoinduced change $\Delta t_\omega$ in the Fresnel coefficient causes a change in the light transmission:

$$S(\omega) = \left(|t_\omega|^2 - |t_\omega + \Delta t_\omega|^2\right)\Big/|t_\omega|^2 \approx -\ 2\ \mathrm{Re}\left(\Delta t_\omega/t_\omega\right). \tag{A9}$$

To estimate $\Delta t_\omega$, we rewrite the right side of eq. (A4) as

$$\Delta\varepsilon(t) \otimes \frac{\partial^2 E^0(t,z)}{\partial t^2} = -\int_{-\infty}^{\infty} d\Omega\; \Omega^2\; \Delta\varepsilon(t) \otimes e^{-i\Omega t}\left(E_\Omega^{0,+} e^{+ik_\Omega z} + E_\Omega^{0,-} e^{-ik_\Omega z}\right). \tag{A10}$$

The Fourier component $\Delta E_\omega(z)$ of $\Delta E(t,z)$ is given by

$$c^2 \frac{\partial^2 \Delta E_\omega(z)}{\partial z^2} + \varepsilon_\omega \omega^2 \Delta E_\omega(z) = -\int_{-\infty}^{\infty} d\Omega\; \Omega^2\; K_{\Omega-\omega}\left(E_\Omega^{0,+} e^{+ik_\Omega z} + E_\Omega^{0,-} e^{-ik_\Omega z}\right), \tag{A11}$$

where the function $K_{\Omega-\omega}$ is given by eq. (9). The general solution of this equation is given by

$$\Delta E_\omega(z) = \int_{-\infty}^{\infty} d\Omega\; \frac{\Omega^2\; K_{\Omega-\omega}}{\Omega^2 \varepsilon_\Omega - \omega^2 \varepsilon_\omega}\; \Delta E_{\omega,\Omega}(z), \tag{A12}$$

where

$$\Delta E_{\omega,\Omega}(z) = E_\Omega^{0,+} e^{+ik_\Omega z} + E_\Omega^{0,-} e^{-ik_\Omega z} + C_\Omega^+ e^{+ik_\omega z} + C_\Omega^- e^{-ik_\omega z}, \tag{A13}$$

and the coefficients $C_\Omega^\pm$ satisfy the boundary conditions:

$$\begin{aligned}
\Delta r_{\omega,\Omega} E_\Omega^i &= \Delta E_{\omega,\Omega}(z=0), & -\Delta r_{\omega,\Omega} E_\Omega^i\; \omega/c &= \Delta E'_{\omega,\Omega}(z=0), \\
\Delta t_{\omega,\Omega} E_\Omega^i &= \Delta E_{\omega,\Omega}(z=d), & \Delta t_{\omega,\Omega} E_\Omega^i\; \omega n_s/c &= \Delta E'_{\omega,\Omega}(z=d),
\end{aligned} \tag{A14}$$

where $\Delta(r,t)_{\omega,\Omega}$ are the corrections to the corresponding Fresnel coefficients. The overall change in the transmission coefficient is given by

$$\Delta t_\omega\; E_\omega^i = \int_{-\infty}^{+\infty} d\Omega\; \frac{\Omega^2\; K_{\Omega-\omega}}{\Omega^2 \varepsilon_\Omega - \omega^2 \varepsilon_\omega}\; \Delta t_{\omega,\Omega}\; E_\Omega^i. \tag{A15}$$

Eqs. (A14) may be combined with eqs. (A6) to obtain, after some algebra,

$$\Delta t_{\omega,\Omega} = t_\Omega - t_\omega + \frac{k_\Omega - k_\omega}{2k_\omega} t_\omega \left\{ r_\Omega - 1 + n_s t_\Omega \left(\cos\delta_\omega - \frac{i}{n_\omega}\sin\delta_\omega\right)\right\}. \tag{A16}$$

Assuming that $\Delta\varepsilon_\omega(t)$ has a time-independent spectral profile (so that $K_{\Omega-\omega} \equiv \Delta\varepsilon_\omega\, K'_{\Omega-\omega}$, see eq. (10)) and $\Omega$ is sufficiently close to $\omega$, we can approximate

$$\frac{\Omega^2\, K_{\Omega-\omega}}{\Omega^2\varepsilon_\Omega - \omega^2\varepsilon_\omega} \approx \frac{\Delta k_\omega}{k_\Omega - k_\omega}\, K'_{\Omega-\omega} \tag{A17}$$

to obtain the final result:

$$S(\omega) = -2\,\mathrm{Re}\left\{\Delta k_\omega \int_{-\infty}^{+\infty} d\Omega\, K'_{\Omega-\omega}\, \frac{E^i_\Omega}{E^i_\omega}\, \Theta_\omega(\Omega)\right\}, \tag{A18}$$

where

$$\Theta_\omega(\Omega) = \frac{t_\Omega/t_\omega - 1}{k_\Omega - k_\omega} + \frac{1}{2k_\omega}\left\{r_\Omega - 1 + n_s t_\Omega\left(\cos\delta_\omega - \frac{i}{n_\omega}\sin\delta_\omega\right)\right\}. \tag{A19}$$

Formula (A18) resembles eq. (12) obtained in Sec. II. In particular, it is easy to demonstrate, by direct calculation, that

$$\Delta k_\omega\, \Theta_\omega(\omega) = \frac{1}{t_\omega}\left(\frac{\partial t}{\partial\omega}\bigg/\frac{dk}{d\omega} + \frac{\partial t}{\partial k}\right)_\omega \Delta k_\omega = \frac{1}{t_\omega}\left(\frac{dt}{d\varepsilon}\right)_\omega \Delta\varepsilon_\omega. \tag{A20}$$

Thus, if the sample is sufficiently thin, so that $n_\omega d/c \ll \tau_p$ (i.e., when $\Theta_\omega(\Omega)$ is a relatively slow function of $\Omega$ over the spectral range of interest), we obtain

$$S(\omega) \approx -2\,\mathrm{Re}\left\{\frac{1}{t_\omega}\left(\frac{dt}{d\varepsilon}\right)_\omega \Delta\varepsilon_\omega \int_{-\infty}^{+\infty} d\Omega\, K'_{\Omega-\omega}\, \frac{E^i_\Omega}{E^i_\omega}\right\}. \tag{A21}$$

For single exponential kinetics, the integral in eq. (A21) is equal to the function $\Phi(\alpha,\beta,\gamma)$ given by eq. (14). In particular, for long group delays $T_e$, the integral in eq. (A21) asymptotically approaches $\exp(-\gamma T_e)$. This applies to the samples of any thickness, since for long group delay times $T_e$, a very narrow interval of frequencies $\Omega$ close to $\omega$ contributes to the integral in eq. (A18)). For a very thin sample ($\delta_\omega << 1$), $t_\omega \approx 2/(n_s + 1)$ and $t_\omega^{-1}(dt/d\varepsilon)_\omega \approx it_\omega \omega d/2c$, so that for $n_s = 1$, eq. (12) is obtained. In the general case, formula (A21) is incorrect and eqs. (A18) and (A19) must be used instead. The integration can be preformed numerically or by expansion of the Fresnel coefficients given by eq. (A7) (that are periodic functions of $\omega$) into a truncated Fourier series and integrating each term analytically, using a modified eq. (14). In most experimental situations, the second term in eq. (A19) is 1-2 orders of magnitude smaller than the first term and may be neglected. Note that the expression for $S(\omega)$ given by eq. (A18) is for *infinite* spectral resolution, i.e., for $\delta = 0$. For $\delta\tau_p << 1$ (which is always the case experimentally), $S(\omega) \approx S(\omega)|_{\delta=0} \otimes g(\omega)$. The latter convolution may be carried out numerically.

# References

\* To whom correspondence should be addressed: *Tel* 630-2529516, *FAX* 630-2524993, *e-mail:* shkrob@anl.gov.

b) Chemistry Division , Argonne National Laboratory,  Argonne, IL 60439

c) CEA/Saclay, DSM/DRECAM/SCM/URA 331 CNRS 91191 Gif-Sur-Yvette Cedex, France

d) See EPAPS Document No. ------------- for the Supplement containing Appendix B and Figs. 1S to 3S with captions (PDF format). This document may be retrieved via the EPAPS homepage (http://www.aip.org/pubservs/epaps.html) or from ftp.aip.org in the directory /epaps/. See the EPAPS homepage for more information.

1. R. A. Crowell, D. J. Gosztola, I. A. Shkrob, D. A. Oulianov, C. D. Jonah, and T. Rajh, Radiat. Phys. Chem., *in press*; N. Saleh, K. Flipo, K. Nemoto, D. Umstadter, R. A. Crowell, C. D. Jonah, and A. D. Trifunac, Rev. Sci. Instrum. **71**, 2305 (2000)

2. M. M. Malley and P. M. Rentzepis, Chem. Phys. Lett. **3**, 534 (1969); Chem. Phys. Lett. **7**, 57 (1970)

3. M. R. Topp, P. M. Rentzepis, and R. P. Jones, Chem. Phys. Lett. **9**, 1 (1971)

4. L. Dhar, J. T. Fourkas, and K. Nelson, Opt. Lett. **19**, 643 (1994)

5. J. T. Fourkas, L. Dhar, K. Nelson, and R. Trebino, J. Opt. Soc. Am. B **12**, 155 (1995); J. T. Fourkas, L. Dhar, and K. A. Nelson, Springer Ser. Chem. Phys. **60**, 141 (1994)

6. G. P. Wakeham, D. D. Chung, and K. A. Nelson, Thermochimica Acta **384**, 7 (2002); W. Wang, D. D. Chung, J. T. Fourkas, L. Dhar, and K. A. Nelson, J. Phys. IV **5-C4**, 289 (1995)

7. A. Brun, P. Georges, G. Le Saux, and F. Salin, J. Phys. D: Appl. Phys. **24**, 1225 (1991)

8. R. Heinicke and J. Grotemeyer, Appl. Phys. B **71**, 419 (2000)

9. S. Backus, C. G. Durfee III, M. M. Murnane, and H. C. Kapteyn, Rev. Sci. Instrum. 69, 1207 (1998)

10. G. S. Beddard, G. G. McFadyen, G. D. Reid, and J. R. G. Thorne, Chem. Phys. Lett. **198**, 641 (1986)

11. J. W. G. Tisch, D. D. Meyerhofer, T. Ditmire, N. Hay, M. B. Mason, and M. H. R. Hutchinson, Phys. Rev. Lett. **80**, 1204 (1998)

12. K. W. DeLong and J. Yumoto, J. Opt. Soc. Am. B **9**, 1593 (1992)

13. I. Wilke, A. M. McLeod, W. A. Gillespie, G. Berden, G. M. H. Knippels, and A. F. G. van der Meer, Phys. Rev. Lett. **88**, 124801 (2002)

14. H. Schulz, P. Zhou, and P. Kohns, "Laser in Forschung und Technik", *Proceedings of the 12$^{th}$ International Congress*, München, June 1995 (Springer, Berlin, 1997); p. 59.

15. (a) S. P. LeBlanc, E. W. Gaul, N. H. Matlis, A. Rundquist, and M. C. Downer, Opt. Lett. **25**, 764 (2000); (b) Z. Jiang and X.-C. Zhang, Appl. Phys. Lett. **72**, 1945 (1998)

16. C. Y. Chien, B. La Fountaine, A. Desparois, Z. Jiang, T. W. Johnston, J. C. Kieffer, H. Pépin, F. Vidal, and H. P. Mercure, Opt. Lett. **25**, 578 (2000); A. Benuzzi-Mounaix, M. Koenig, J. M. Boudenne, T. A. Hall, D. Batani, F. Scianitti, A. Masini, and D. Di Santo, Phys. Rev. E **60**, R2488 (1999); J.-P. Geindre, P. Audebert, S. Rebibo, and J.-C. Gauthier, Optics Lett. 26, 1612 (2001); K. Y. Kim, I. Alexeev, and H. M. Milchberg, Appl. Phys. Lett. 81, 4124 (2002).

17. E. Tokunaga, A. Terasaki, and T. Kobayashi, in "Ultrafast Processes in Chemistry and Photobiology", edited by M. A. El-Sayed, I. Tanaka, and Y. Molin (Blackwell, Oxford, 1995); ch. 10, p. 257.

18. S. A. Kovalenko, A. L. Dobryakov, J. Ruthmann, and N. P. Ernsting, Phys. Rev. A **59**, 2369 (1999) and references therein.

19. J.-K. Wang, T.-L. Chiu, C.-H. Chi, and C.-K. Sun, J. Opt. Soc. Am. B **16**, 651 (1999)

20. K. Duppen, F. de Haan, E. T. J. Nibbering, and D. A. Wiersmaa, Phys. Rev. A **47**, 5120 (1993); A. N. Naumov and A. M. Zheltikov, Laser Phys. **11**, 934 (2001).

21. I. A. Shkrob, D. A. Oulianov, R. A. Crowell, and S. Pommeret, J. Appl. Phys., *in press*.

22. I. A. Shkrob, D. A. Oulianov, and R. A. Crowell, J. Opt. Soc. B, *submitted*.

23. S. Mukamel, *Principles of Nonlinear Optical Spectroscopy* (Oxford University Press, New York, 1995).

24. J. A. Moon and J. Tauc, J. Appl. Phys. **73**, 4571 (1993); D. M. Roberts, J. F. Palmer, and T. Gustafson, J. Appl. Phys. **60**, 1713 (1986).

**Figure captions.**

**Fig. 1**

Real *($\Phi'$)* and imaginary *($\Phi''$)* part of function $\Phi(\alpha,\beta,\gamma)$ introduced in eq. (14) plotted as a function of the frequency offset $\Delta\omega = \omega - \omega_0$ (to the bottom) and the group delay $T_e$ (to the top; eq. (17)). The real part corresponds to the contribution from transient absorption; the imaginary part corresponds to the contribution from nonlinear refraction. The dash-dot bell-like trace is the spectrum of the probe pulse, $|E_\omega|^2$. The dotted trace is the exponential kinetics convoluted with the Gaussian pump and probe pulses. The following parameters were assumed:

$\tau_p = 20\,fs$, $\tau_L = 100\,fs$, $s = 2048$, $\delta = 2\,cm^{-1}$, $T = 30\,ps$, and $\gamma^{-1} = 40\,ps$.

**Fig. 2.**

A closer look at the oscillation pattern for the real and complex parts of function $\Phi$ (eq. (14) for $\tau_p = 20\,fs$, $s = 2048$, and $T = \gamma = 0$. Solid traces are for $\tau_L$=100 fs and $\delta$ =2 cm$^{-1}$; dotted traces are for $\tau_L = \delta = 0$. The group delay is given in the units of $\tau_{GVD}$ (equal to 905 fs). The extrema for $\Phi'$ correspond to the saddle points for $\Phi''$, and *vice versa*. The positions of the extrema are given by eqs. (24), (25), and (26).

**Fig. 3.**

Real ($\Phi'$) part of function $\Phi(\alpha,\beta,\gamma)$ introduced in eq. (14) plotted as a function of the frequency offset $\Delta\omega$ (bottom) and group delay $T_e$. Save for the spectral resolution $\delta$ and pump pulse duration $\tau_L$, the parameters are the same as in Fig. 1. For traces (a-d) $\tau_L$=0 and $\delta$=0 (a), 2 cm$^{-1}$ (b), 5 cm$^{-1}$ (c) and 10 cm$^{-1}$ (d), respectively. For traces (e-h) $\delta$=0 and $\tau_L$=100 fs (e), 300 fs (f), 500 fs (g) and 1 ps (h), respectively.

**Fig. 4.**

Real part of function $\Phi$ (eq. (14)) as a function of the stretch factor: (a) $\gamma^{-1}= 1$ ps, $T=1.25$ ps, and $s=100$ (GVD of $4\times10^4$ fs$^2$), (b) $\gamma^{-1}=5$ ps, $T=6.25$ ps, and $s=500$ (GVD of $2\times10^5$ fs$^2$), and (c) $\gamma^{-1}=50$ ps, $T=62.5$ fs and $s=5000$ (GVD of $2\times10^6$ fs$^2$). Other calculation parameters: $\tau_p=20$ fs, $\tau_L=100$ fs, and $\delta=2$ cm$^{-1}$. The higher the GVD, the smaller the section of the kinetics exhibiting the oscillations.

**Fig. 5.**

Simulated $S(\omega)$ kinetics (plotted against the group delay $T_e$) as a function of phase $\phi_\varepsilon$ of a frequency-independent photoinduced change $\Delta n_\omega$ in the complex refraction index (same parameters as in Fig. 1). This phase (in degrees) is indicated in the plot.

Figure 1; Shkrob and Pommeret

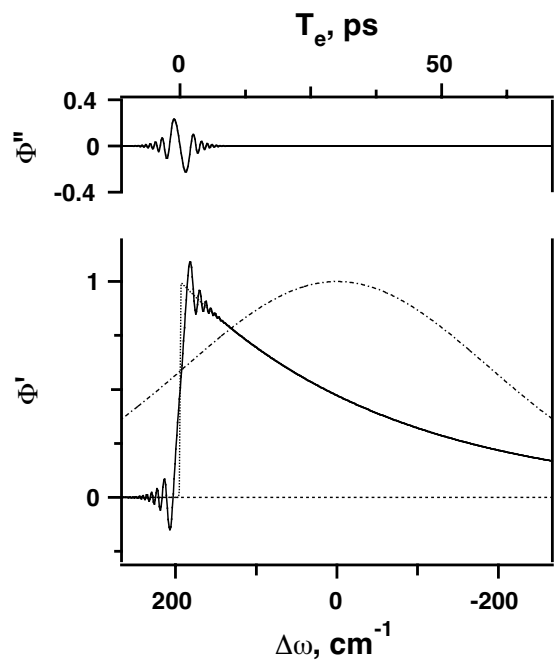

Figure 2; Shkrob and Pommeret

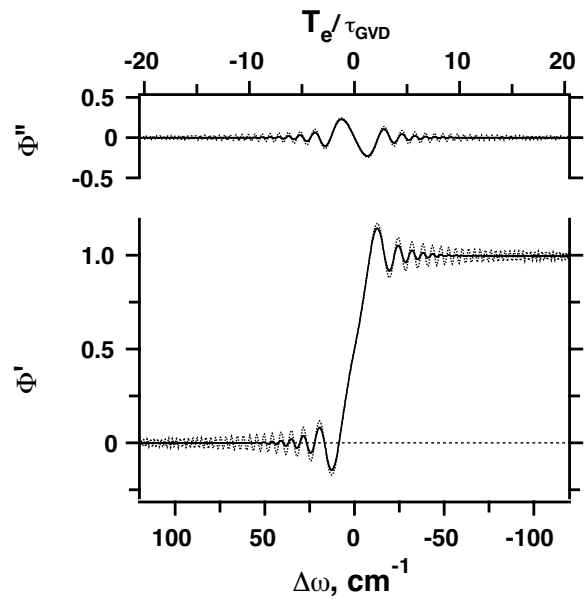

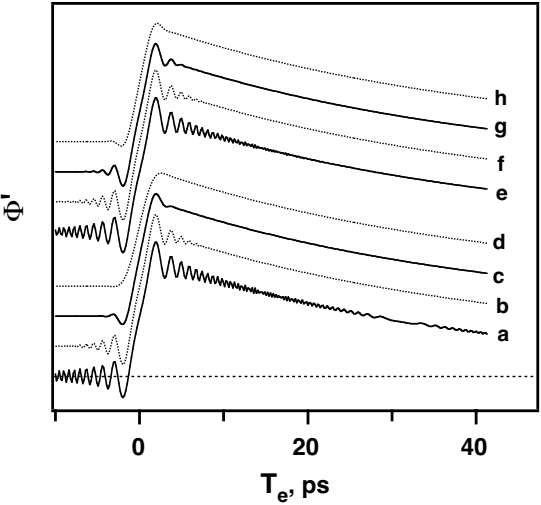

Figure 4; Shkrob and Pommeret

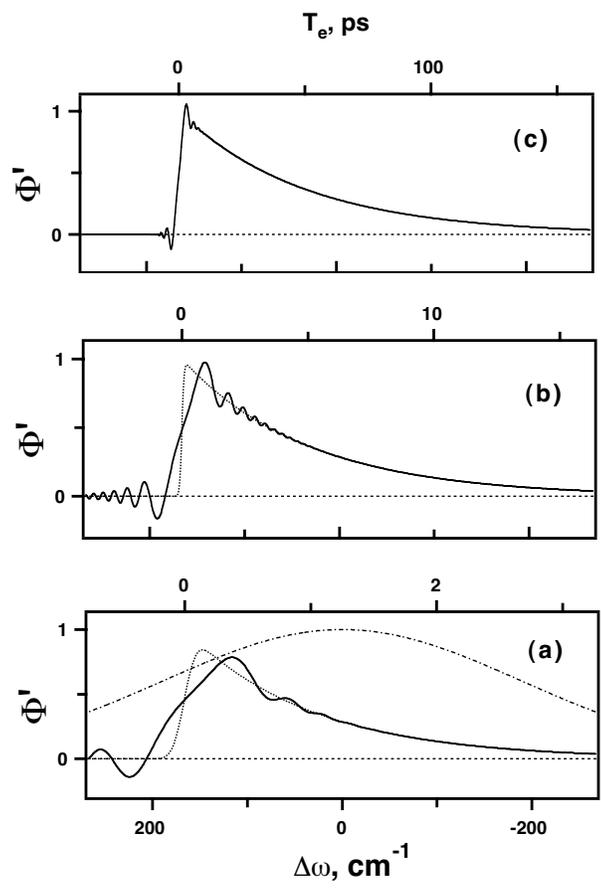

Figure 5; Shkrob and Pommeret

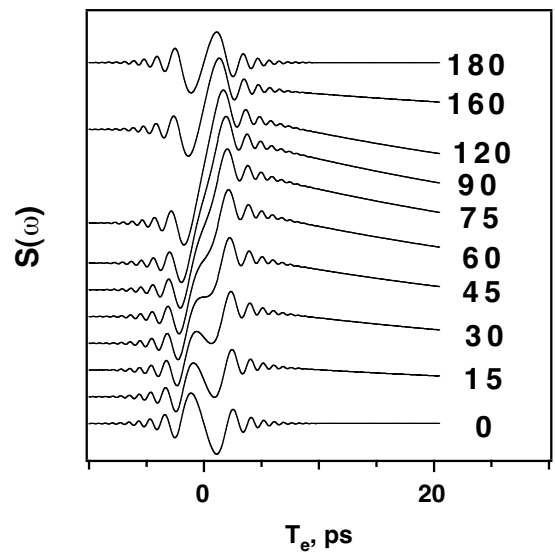

# Theoretical Analysis of Frequency-domain "single-shot" (FDSS) ultrafast spectroscopy.

Ilya A. Shkrob [a)] and Stanislas Pommeret [a),b)]

[a)] *Chemistry Division, Argonne National Laboratory, Argonne, IL 60439*

[b)] *CEA/Saclay, DSM/DRECAM/SCM/URA 331 CNRS 91191 Gif-Sur-Yvette Cedex, France*

shkrob@anl.gov

## Supporting Information.

## Appendix B: Experimental considerations: Analysis.

### 1. Clipping of dispersed pulse.

To demonstrate the invariance of FDSS kinetics to clipping of the probe pulse dispersed on the compressor grating, eq. (12) was used to calculate $S(\omega)$ for a probe pulse whose Gaussian spectrum was truncated at $\Delta\omega \approx \pm 1/\tau_p$ (Fig. 2S). It is clear from this calculation that the effect of truncation is limited to narrow regions near the truncation points. This is explained as follows:

The clipping of the spectrum is equivalent to narrowing the integration interval over $\Omega$ (or $\mu$) in eq. (12). The integral function is damped as $\exp\left(-\mu^2\left\{\tau_L^2/4 + \tau_p^2/2\right\}\right)$, and for $\tau_L$=100 fs pump and $\tau_p$=20 fs probe the Gaussian envelope is 160 cm$^{-1}$ FWHM. If the kinetic origin is removed from the truncation points by that much, the cutoff has no effect on the oscillation pattern. In general, the effect of limiting $\mu$ to a range of $(-M, +M)$ is equivalent to introducing an additional damping factor $\exp(-\mu^2/M^2)$ that increases $\alpha^2$ (given by eqs. (15) and (20)) by $M^{-2}$. For instance, a bandpass of $\pm 10$ cm$^{-1}$ is equivalent to increasing $\tau_L$ to 1 ps. Thus, a few picoseconds away from the kinetic origin, the truncation of the probe pulse has no effect since a very narrow band of the frequencies $\Omega$ contributes to $S(\omega)$ in the first place. Only for very low GVD ($s$ of $\pm(10\text{-}50)$) can the truncation distort the kinetics. However, for those GVD's, the spread of the beam on the gratings is small and no clipping occurs anyway. This consideration indicates that once the kinetic origin is removed by 100-200 cm$^{-1}$ from the clipping points, the kinetics (including the oscillation pattern) do not change at all.

Concerning the dispersion achievable using the standard CPA optics, a $\tau_p$=20 fs (33 fs FWHM), 800 nm pulse from a Ti:sapphire oscillator yields an optimum sampling interval (eq. (29)) of $\pm 250$ cm$^{-1}$. Simple estimates show that for the standard 1200 g/mm ($\theta_L$= 28.7°), 11 cm - wide reflection grating used in the first order, the distance $L_g$ can be increased to 1.5-2 m without adverse effects on the FDSS kinetics: For $\theta$=40° and $L_g$=1 m, $\phi''(\omega_0)$=-3 ps$^2$ ($s$=-7650, $\xi_3$=1.34), and the $\pm 250$ cm$^{-1}$ interval would be dispersed over 4.5 cm.

The use of smaller stretch factors ($<10^3$) and longer probe pulses ($>$ 200 fs) might be preferable to compressing a 10-50 fs pulse by a large factor of -(1-10)x10$^3$, as done in our experiments.[21] However, since the oscillations take $\approx 1/\sqrt{s}$ of the total kinetics regardless of the probe pulse width, larger GVD and shorter pulses actually yield "cleaner" TA traces, even for relatively long sampling times. Furthermore, resolving the oscillation pattern requires short pump pulses and, therefore, short probe pulses. In the end, the only advantage of longer probe pulses is the possibility of stretching these pulses with optical fibers.

Since it is impossible to place large reflection gratings (needed to obtain large GVD) closer than 15-20 cm, due to the beam geometry, an equivalent positive chirp has to be introduced prior to the compression, as done in ref. 21. Alternative schemes for obtaining small stretch/compression factors ($10^2$-$10^3$) (other than slight overcompression of stretched pulses, as done in ref. 21) may include the use of transmission gratings (that can be put close to each other and, thereby, give the widest dynamic range),[a] dispersive materials (such as SF18 glass),[9] and optical fibers (especially for visible light).[10] For instance, SF18 glass has GVD of 1540 fs$^2$/cm at 800 nm ($s$=+3.85/cm, $\xi_3$=0.5) [9] and a stretch factor of +100 can be achieved by passing a transform limited probe pulse through 25 cm of this material. Beddard et al. [10] used a 2.5 m long single mode optical fiber to obtain a GVD of 0.14 ps$^2$ to perform their first FDSS experiment. Thus, one can envision a hybrid system where low (positive) stretch factors are obtained by using dispersive materials whereas large (negative) compression factors are obtained using compressors with a reflection grating.

---

[a]  J.-K. Rhee, T. S. Sosnowski, T. B. Norris, J. A. Arns, and W. S. Colburn, Opt. Lett. **19**, 1550 (1994)

## 2. Probe wavelength tuning.

The most popular method for varying the probe wavelength in PPS experiments is by generation of a white light supercontinuum and using narrow band interference filters to select a wavelength range. The supercontinuum is nonlinearly chirped, with a typical stretch factor of 20-100. Due to large beam divergence, it is difficult to compress these pulses using a gratings compressor (although, low GVD can be obtained by dispersion in a fiber).[10,18] Thus, one is forced to use optical parametric amplifiers (OPA's) as the source of coherent probe light. Furthermore, the compressor must be realigned for each probe color (due to the need to keep the incident angle reasonably close to the Littrow angle). While this operation can be automated, it is difficult to avoid large changes in the stretch factor since $s \propto \lambda^3$ (eq. (30)). A change-over from 800 nm to 400 nm decreases GVD by an order of magnitude. With the exception of low GVD's, this decrease cannot be compensated by the equivalent increase in $L_g$. Also, for $\lambda$ other than the oscillator wavelength there is no built-in source of stretched probe pulses, and the dynamic range is reduced.

It should be stressed that most of these problems present themselves for shorter wavelengths; for longer wavelengths, the change-over is less problematic. Even for shorter wavelengths, a ready solution exists provided that the probe beam is sufficiently bright to tolerate losses that occur when the gratings are used in higher diffraction orders. A typical OPA yields 2-10 µJ of the visible light, whereas less than 1 nJ is needed to probe the sample. For a given groove spacing, the number of possible orders $m$ (given by $\lambda/d_g < 2$) increases as $1/\lambda$ whereas GVD increases as $m^2$, and the loss of GVD at the shorter wavelengths can be compensated by increasing *m*. E. g., for the standard 1200 *g*/mm grating compressor, 4 diffraction orders exist at 400 nm, and the Littrow angle for the third order is still acute, ca. 46º. With this grating used in the 2$^{nd}$ and 3$^{d}$ orders, the GVD at 400 nm is 0.5 and 2.3 times that at 800 nm in the first order. The chart shown in Fig. 3S helps to choose the optimum $L_g$ and *m* to obtain a desirable stretch factor *s* for a given wavelength *λ*, for a 33 fs FWHM pulse and 1200 *g*/mm grating (of the fixed width).

**3. The spectrum of the probe pulse.**

While the theoretical treatment of section II is formulated for an ideal Gaussian pulse, the spectra of probe pulses obtained using CPA (before or after the compression) often exhibit low-amplitude "ripple" (2-10%) juxtaposed on the Gaussian profile. The origin of this "ripple" can be traced to etalon and aperture effects during the passage of the beam through the amplification optics and to the modes of Ti:sapphire oscillator itself. While this "ripple" is reduced when the beam passes through a compressor, it reappears as the probe light is diffracted by the sample and the focussing optics. Due to this diffraction, the amplitude of the "ripple" is not uniform across the beam; the central part exhibits 2-3 times stronger oscillations than the outer part. A typical slit opening of the monochromator needed to obtain 1-5 cm$^{-1}$ resolution is 50-100 μm and, in the absence of a diffuser, it is the central part of the beam that is imaged on the detector. Since the reference and signal beams travel different paths and are diffracted differently, their "ripple" patterns are not identical. While most of the "ripple" is divided out by the normalization of the signal spectrum by the reference spectrum,[21] better results are obtained when a diffuser is inserted to homogenize the light. Though the "ripple" pattern is still observed, close semblance between the "ripple" patterns for the signal and reference beams is achieved.

While this "ripple" is a nuisance, it has almost no effect on the FDSS kinetics. To a first approximation, the "ripple" can be simulated as

$$E_\omega = E_\omega^0 \left\{1 + A_{osc} \, \exp\left(i[\omega - \omega_0]/\omega_{osc}\right)\right\} \qquad \text{(B1)}$$

where $E_\omega^0$ is the $\omega$ component of the field without the "ripple", $A_{osc}$ (<<1) is the complex amplitude of the oscillations, and $\omega_{osc}$ is their period. For the field given by eq. (B1), the ratio in eq. (12) is given by

$$E_{\omega+\mu}/E_\omega \approx E_{\omega+\mu}^0/E_\omega^0 \left\{1 + A_{osc} \exp(i\mu/\omega_{osc})\right\} \qquad (B2)$$

and

$$\begin{aligned}S(\omega) &\approx \frac{2\omega d}{c} \mathrm{Im}\, \Delta n_\omega \{\Phi(\alpha,\beta,\gamma) + A_{osc}\Phi(\alpha,\beta - i\omega_{osc}^{-1},\gamma)\} \approx \\ &\frac{2\omega d}{c}\mathrm{Im}\,\Delta n_\omega [1 + A_{osc}]\, \Phi(\alpha,\beta,\gamma)\end{aligned} \qquad (B3)$$

The last approximation in eq. (B3) is justified by the fact that offsetting the coefficient $\beta$ is equivalent to a shift of the delay time $T$ of the pump by $\omega_{osc}^{-1}$ - which is a small correction relative to the characteristic time $\tau_{GVD}$ of the oscillations shown in Figs. 1 to 5. We conclude that the high-frequency "ripple" is inconsequential; at most, it adds a small shift to the phase $\phi_\varepsilon$ of $\Delta n_\omega$.

# Figure captions (1S to 3S)

**Fig. 1S.**

Effect of nonzero third order dispersion on the frequency-domain kinetics $S(\omega)$ for pure photoabsorption ($\phi_\varepsilon=90^\circ$). Same parameters as in Fig. 1 except for $\delta = 0$. The dashed trace for $\xi_3 = 0$ was obtained analytically using eq. (14); the solid trace for $\xi_3/\omega_0\tau_p = -0.034$ (calculated using eq. (31) for Littrow angle dispersion of 800 nm pulse on a 1200 $g$/mm grating in the first order) was obtained by numerical integration of eq. (12). In (a), both traces are plotted as a function of the group delay $T_e$ calculated for GVD only (eq. (17)). In (b) the trace for nonzero $\phi'''(\omega_0)$ is plotted vs. the group delay time $T_e$ given by eq. (32).

**Fig. 2S.**

Effect of truncation of the probe spectrum on the $S(\omega)$ kinetics (same parameters as Fig. 1 except for $\delta = 0$). The dashed curve shows the clipped wings of the probe pulse spectrum. The bold line indicates the $|\Delta\omega|<250$ cm$^{-1}$ section. The thin solid line is the kinetics obtained with the truncated pulse (by numerical integration of eq. (12)); the dotted line is the same kinetics obtained with the full spectrum. The two kinetic traces are the same except for the two narrow regions at the truncation points.

**Fig. 3S.**

Center wavelength *(λ)* dependence of (a) the Littrow angle *(dotted lines)* and the beam spread (thick solid lines) on the grating and (b) stretch factor (*thick solid lines*; negative for a compressor) and third-order dispersion *(dotted lines)* for $\tau_p$=20 fs probe pulse ($|\Delta\omega|<267$ cm$^{-1}$ "optimum range") dispersed using a 1200 $g$/mm grating in the *m*-th diffraction order at $L_g$=100 cm (the GVD and the beam spread are proportional to this length). The third order dispersion (eq. (31)) is given as $-\xi_3/\omega_0\tau_p$, where $\omega_0 = 2\pi\lambda/c$.

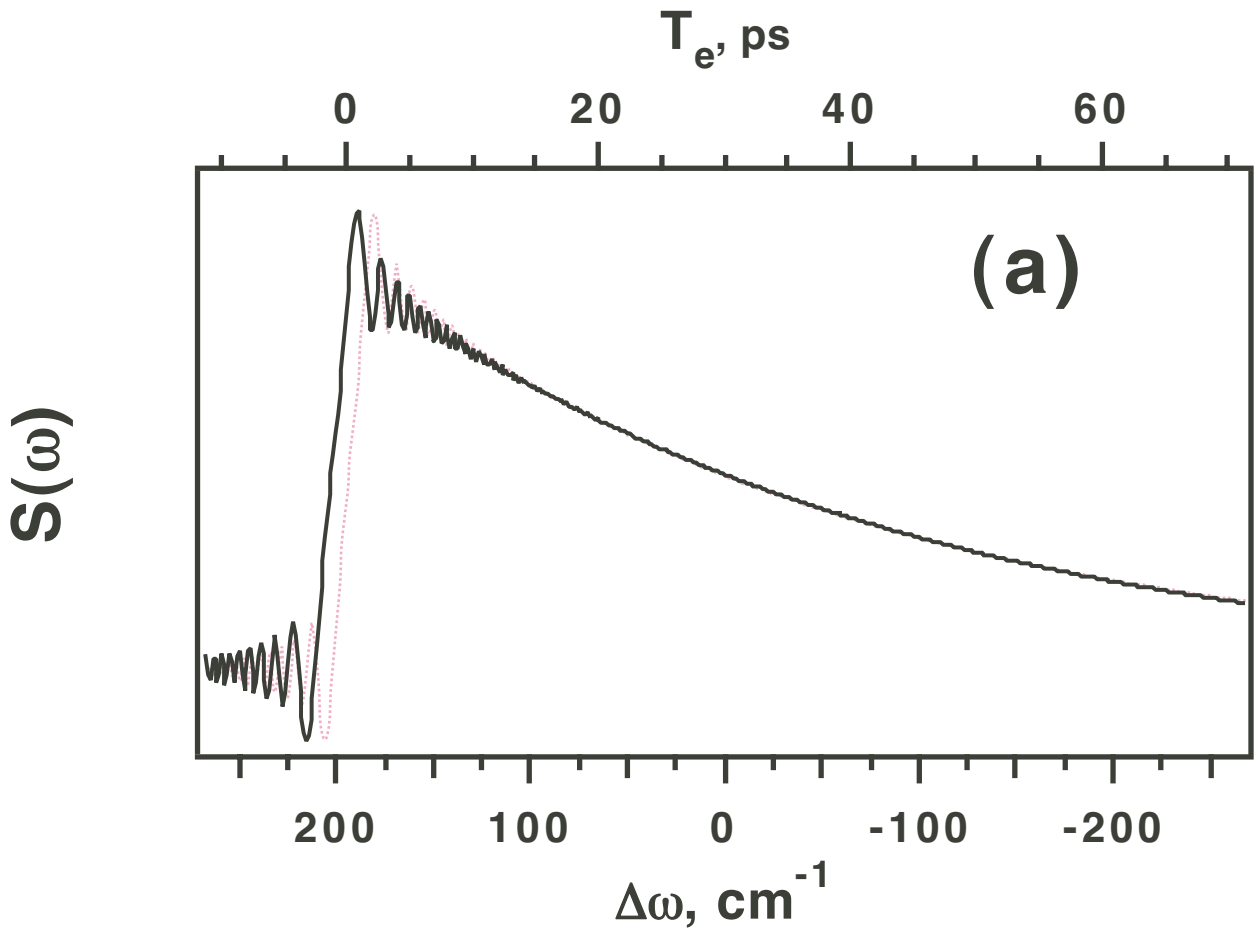
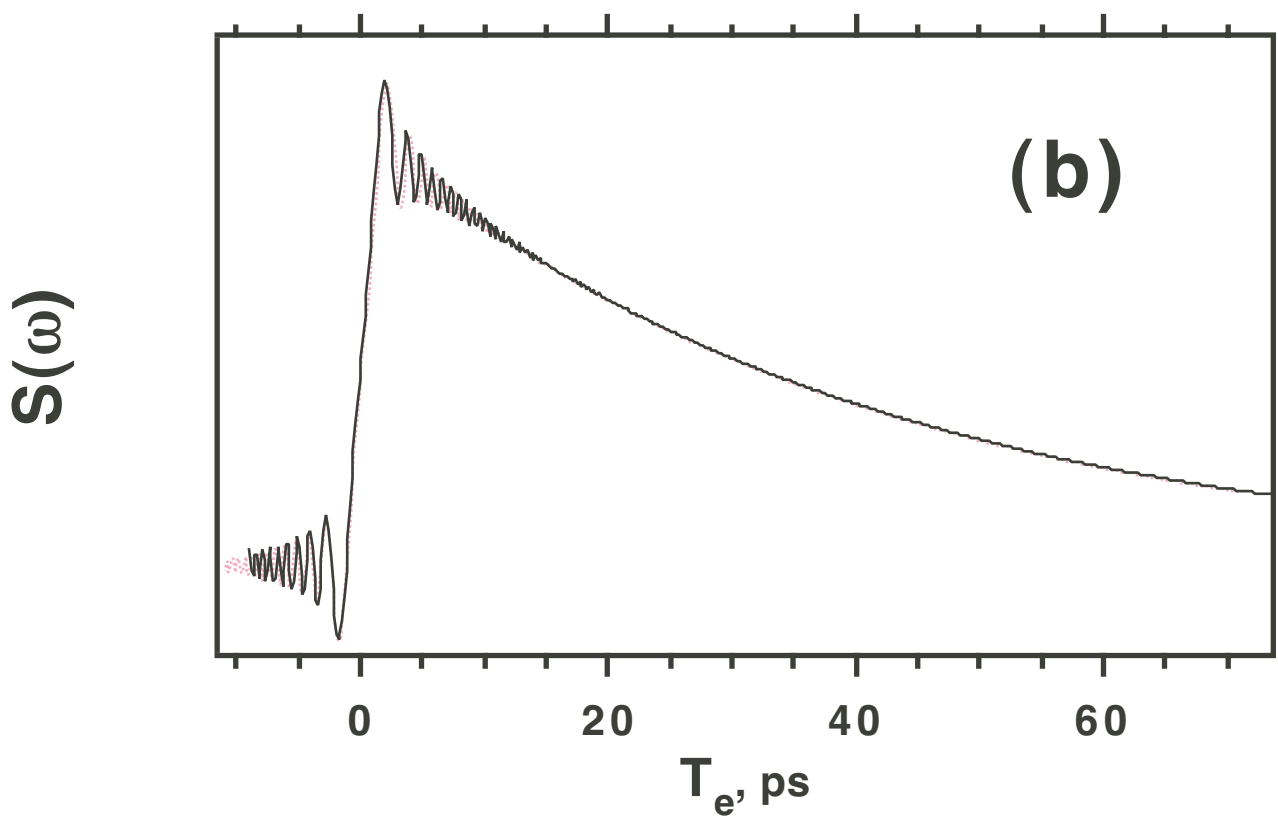

**Fig. 1S; Shkrob**

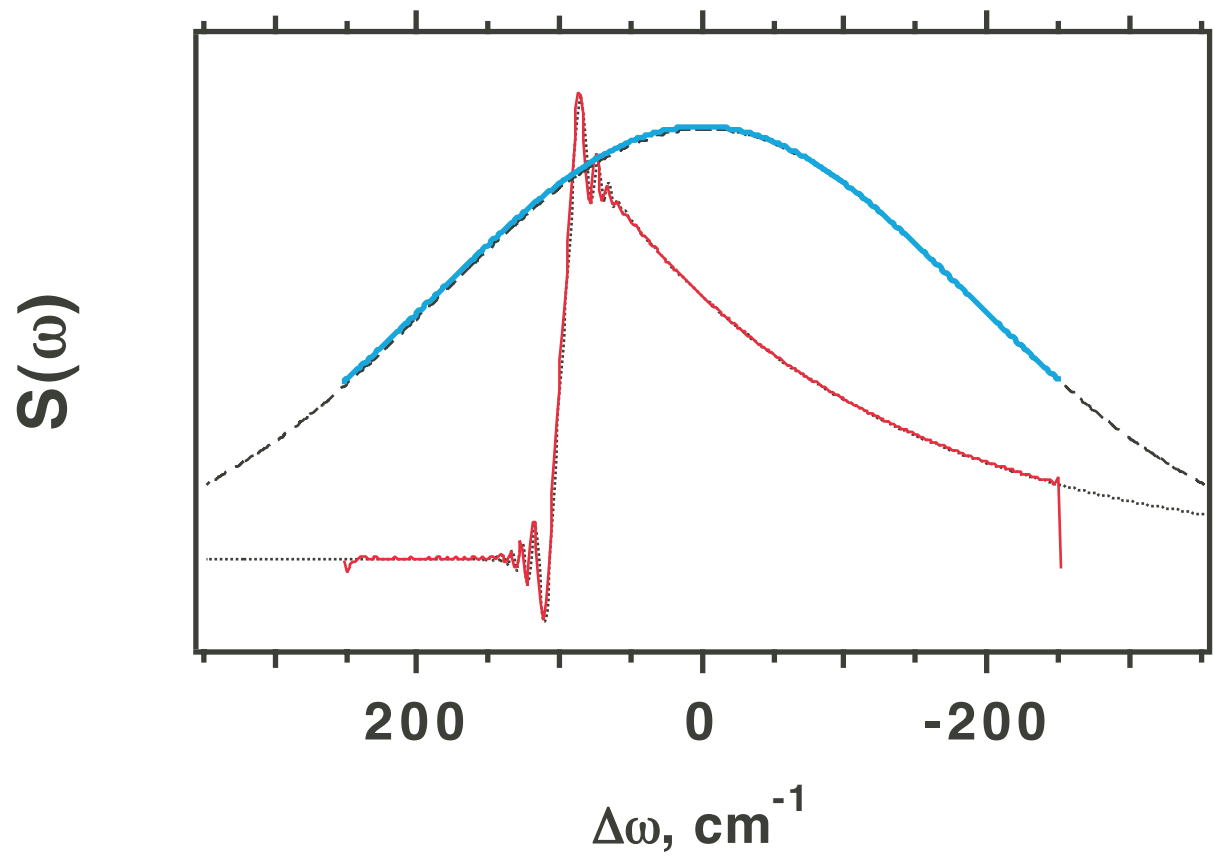

Fig. 2S; Shkrob

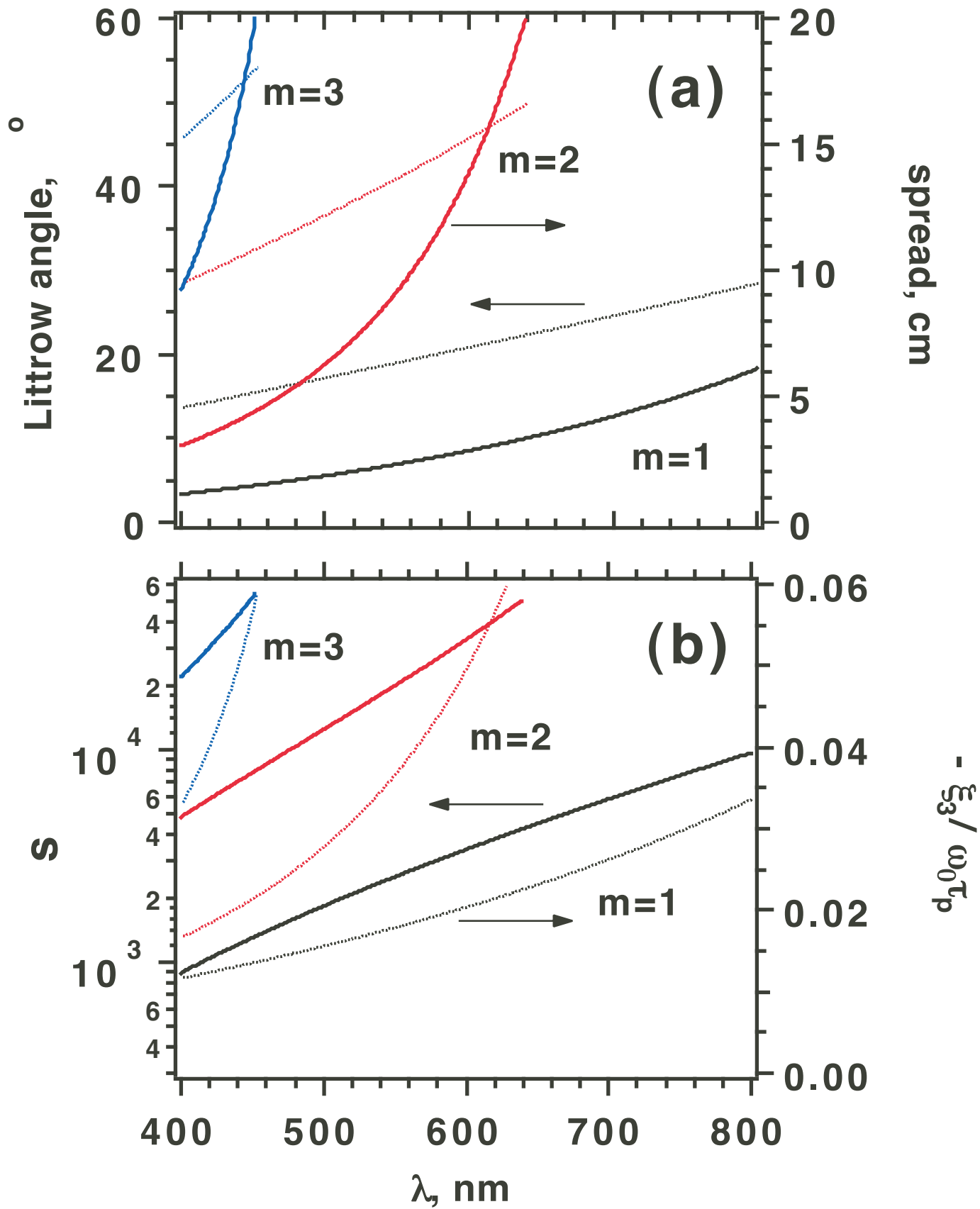

Fig. 3S; Shkrob